# Thermal Conductivity Measurement Using Modulated Photothermal Radiometry for Nitrate and Chloride Molten Salts

Ka Man Chung[1], Tianshi Feng[2], Jian Zeng[2], Sarath Reddy Adapa[2], Xintong Zhang[2], Andrew Z. Zhao[1], Ye Zhang[3], Peiwen Li[3], Youyang Zhao[4], Javier E. Garay[1, 2], Renkun Chen[#1.2]

[1]Program in Materials Science and Engineering, University of California, San Diego, La Jolla, California 92093, United States

[2]Department of Mechanical and Aerospace Engineering, University of California, San Diego, La Jolla, California 92093, United States

[3]Department of Aerospace and Mechanical Engineering, The University of Arizona, Tucson, AZ 85721, United States

[4]National Renewable Energy Laboratory, 15013 Denver West Parkway, Golden, CO 80401, United States

#Corresponding author: rkchen@ucsd.edu

## Abstract

Molten salts are being used or explored for thermal energy storage and conversion systems in concentrating solar power and nuclear power plants. Thermal conductivity of molten salts is an important thermophysical property dictating the performance and cost of these systems, but its accurate measurement has been challenging, as evidenced by wide scattering of existing data in literature. The corrosive and conducting nature of these fluids also leads to time consuming sample preparation processes of many contact-based measurements. Here, we report the measurement of thermal conductivity of molten salts using a modulated photothermal radiometry (MPR) technique, which is a laser-based, non-contact, frequency-domain method adopted for molten salts for the first time. By unitizing the advantages of front side sensing of frequency-domain measurements and the vertical holder orientation, the technique can minimize the natural convection and salt creeping effects, thus yielding accurate molten salt thermal conductivity. The MPR technique is first calibrated using standard molten materials including paraffin wax and sulfur. It is then applied on measuring pure nitrate salts ($NaNO_3$ and $KNO_3$), solar salt ($NaNO_3$–$KNO_3$ mixture), and chloride salt (NaCl–KCl–$MgCl_2$). The measurement results are compared with data from literature, especially those obtained from laser flash analysis (LFA). Our results demonstrate that the MPR is a convenient and reliable technique of measuring thermal conductivity of molten salts. Accurate thermal conductivity data of molten salts will be valuable in developing the next-generation high-temperature thermal energy storage and conversion systems.

## 1. Introduction

Molten salts are an important class of heat transfer fluids (HTFs) in high-temperature thermal energy systems including concentrating solar power (CSP), nuclear power plants, and thermal energy storage (TES). Solar salt ($NaNO_3$–$KNO_3$ mixture with wt.% 60–40) is being used in the state-of-the-art CSP plants with TES capability at temperatures up to 550 $^oC$ [1, 2]. Eutectic chloride salts with stability above 800 $^oC$ are being evaluated for next-generation CSP plants operating at high temperatures for higher energy conversion efficiency [3-5]. In these cases, molten salts are used as both the HTFs collecting the concentrated solar heat and as high-temperature TES media. Molten fluoride salts have gained great attention in the development of generation-IV nuclear power plants as nuclear reactor coolants [6, 7]. These salts, such as LiF–$BeF_2$ and LiF–NaF–KF, have high thermal stability, satisfactory specific heat and thermal conductivity at temperatures higher than 700 $^{o}$C [8, 9], thus making them well suited for molten salt reactors (MSRs).

Thermal conductivity of molten salts is important for the performance and cost of these systems. For instance, the thermal conductivity of molten salts dictates the size and thus the cost of molten salt-steam heat exchangers in a CSP power plant. However, reliable thermal conductivity data of molten salts are difficult to obtain. Even common nitrate salts have shown large discrepancies as reported in the literature. For example, there is around ~40 % difference among the reported thermal conductivity values for the solar salt [10], despite its widespread use in commercial CSP plants. Reliable data on the emerging high temperature chloride salts are even more scarce. Molten salts are difficult to measure using conventional contact methods because of their corrosive and conductive nature [11, 12]. In the commonly used transient hot-wire (THW) method, an electrically insulated and corrosion-resistant coating is needed for metallic hot wires [13, 14], which complicates the sample preparation process. The corrosion-resistant coating can only slow down and minimize the corrosion effect. However, the hot wires and sensors would eventually be damaged, leading to measurement errors. In steady state methods, natural convection could introduce substantial errors when there is a large molten salt volume contained in test sections [11, 12, 15-17]. The transient thermal signals in both the steady state and THW methods could also be highly influenced by the ambient temperature fluctuations. Radiative heat loss could affect the thermal conductivity measurements in the steady state techniques as well [18]. Recently, laser flash analysis (LFA) is becoming increasingly popular for molten salts measurements due to its non-contact nature and ease of operation. However, LFA data on molten salts have also witnessed large scattering, due to many factors including the salt creeping effect [19, 20]. When the creeping occurs, salts tend to form precipitates on the wall of crucibles used to hold the molten salts, leading to uneven thickness of the molten salt layer in the crucible and large uncertainty in the LFA data. The time domain nature of the LFA also makes it susceptible to convection and radiation heat loss from the crucible at high temperature, especially if the time window used for data fitting is not properly chosen [21-24]. Zhao *et al.* [12] recently developed a frequency-domain hot-wire technique for measuring molten nitrate salts. The thermal penetration depth of the modulated heating power on the hot wires is confined to about 100 μm within the molten salt volume to minimize the convection and other heat loss effects. Their thermal conductivity results of molten nitrate salts are ∼11 to 15 % higher than the existing reference values obtained from conventional steady state and time domain measurements, including the THW and LFA methods. Their results show more reasonable temperature dependence of thermal conductivity compared to previously reported values and are consistent with their predictions by a phonon gas model for dense and strongly interacting liquids [25]. This study demonstrates the value of frequency-domain technique for obtaining accurate results. Nevertheless, the use of a hot wire makes it less convenient, especially for corrosive high-temperature chloride salts.

Recently, we have developed a non-contact modulated photothermal radiometry (MPR) technique to measure bulk solids and coatings at high temperatures [26, 27]. It is a front side sensing photothermal radiometry in which the oscillating thermal emission induced by an intensity-modulated heating laser is utilized for thermometry. The technique was further applied to the measurements of stationary and flowing HTFs such as water and thermal oil up to ~170 $^{o}$C [27]. In this work, it is applied for molten salt measurements for the first time. A special molten salt holder with a vertical configuration is designed for the MPR setup to address several issues in the LFA, in particular, the salt creeping effect. Measurement errors due to the natural convection effect can be minimized in the MPR method by carefully controlling thermal penetration depth into the molten salt layers. With this new technique, the thermal conductivity of molten nitrate and chloride salts up to 700 $^{o}$C, with less than 10 % uncertainty is obtained. The selected molten salts for measurements include pure nitrate salts ($NaNO_3$ and $KNO_3$), solar salt ($NaNO_3$–$KNO_3$ mixture) and ternary chloride salt (NaCl–KCl–$MgCl_2$). The measured data are also compared with the values from the LFA, and other techniques reported in the literature. Our results demonstrate the reliability of the MPR method for measuring thermal conductivity of molten salts. Therefore, the MPR technique can serve as a useful and convenient tool to obtain accurate thermophysical properties of molten salts.

## 2. Experimental Procedure

### 2.1. MPR setup

Figure 1 shows the principle of the MPR measurement and the schematic of the molten salt holder in the MPR setup. The photographs of the setup are shown in Supplementary Figure S1. A continuous-wave laser with its intensity modulated at an angular frequency $\omega$ ($q(t) = q_s e^{j\omega t}$) was heating the front surface of the sample holder. The temperature response of the same surface was measured using a high-speed infrared (IR) detector. The MPR setup was the same as in our earlier work [26, 27]. The laser profile was converted from Gaussian to a top-hat (uniform) using a homogenizer. The diameter of the uniform intensity portion of the laser beam was about 15 mm. In this work, we used either a liquid-nitrogen-cooled Hg-Cd-Te (MCT) detector (Kolmar Technologies, Inc., KMPV11-0.25-J1/DC70, 62 ns response time) or an uncooled PbSe detector (Thorlabs, PDA20H-EC, 34 ms response time). The IR detection area was about 0.5 mm and 2 mm respectively for these two detectors. The voltage reading from either detector was converted to temperature through a calibration using a pyrometer (Lumasense Technologies IGA 320/23-LO). We confirmed that both detectors yielded equivalent results when using their respective calibration curves. The heat flux $q_s$ in the measurements was calibrated by using a known reference sample. Borosilicate glass was chosen as the reference sample because it has well-known and stable thermophysical properties [28, 29]. The $c_p$ of molten salts was additionally verified by us using differential scanning calorimetry (DSC). For each temperature point, the measurement took about 20 minutes, as a sufficiently long data collection time was used to reduce the noise at each frequency point within the frequency range (e.g., 0.1 to 10 Hz).

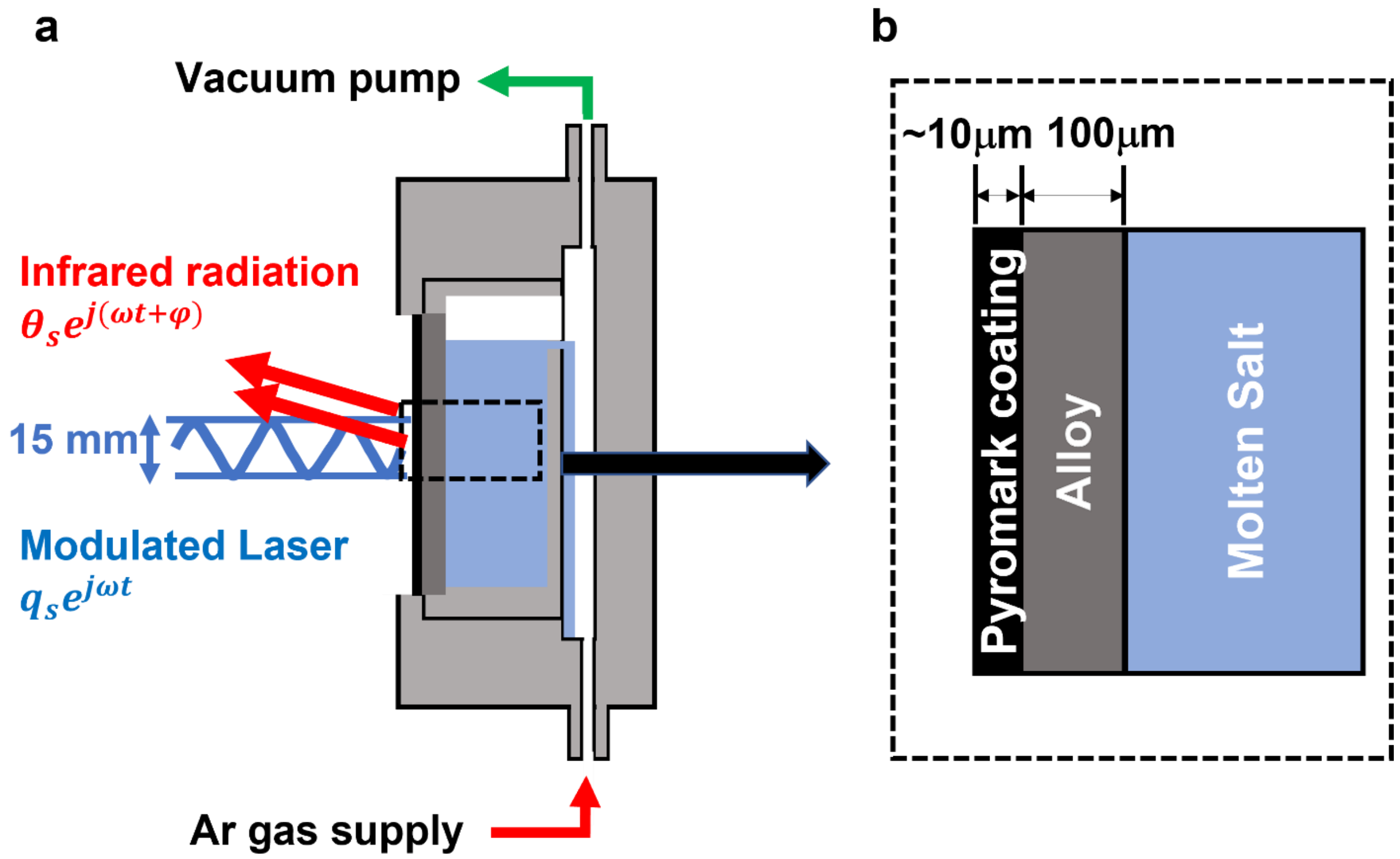

**Figure 1.** Schematics of the MPR molten salt measurement. **(a)** In the MPR measurement, the molten salt holder is vertically positioned. The thermal response from the front side of the molten salt holder is detected. The small cavity on the backside of the holder allows the overflow of the molten salt; **(b)** the three-layer structure formed by the molten salt, Inconel 625 cover and the Pyromark coating used for the heat transfer model in MPR.

A special MPR holder was custom-made for molten salt measurements at high temperatures (up to ~700 $^{o}$C). The design is beneficial for front side sensing and vertical configuration. The vertical configuration can eliminate the salt creeping effect which is one of the challenging issues commonly observed in molten salt thermophysical property measurement. A detailed discussion of the benefits of the molten salt holder design and the comparison of the MPR method to the other measurement techniques will be discussed in *section 3.6*. The holder contained three main components: an Inconel 625 backplate with a small cavity on the front side and a larger cavity on the backside, a thin (100 μm thick) Inconel 625 sheet covering the salt, and an Inconel front plate with a middle opening that secured the Inconel 625 cover to the Inconel backplate with screws. The front cavity within the backplate had a depth of ~3 mm and a diameter of 50 mm and it was used for holding the salt during the measurements. Inconel was selected as the holder material because it is a corrosion-resistant containment material. The back cavity in the backplate was connected to Argon (Ar) gas flow and a vacuum pump. There was a small hole around the top edge connecting the two cavities, which allowed the gas to flow to the front cavity and overflow of molten salt to the back cavity. Purging Ar gas throughout the measurement enabled the salt samples to be free of oxygen and moisture contamination. The backplate contained inserted cartridge heaters to reach the measurement temperature. The front plate had an opening of ~40 mm-diameter in the middle for optical access to the Inconel 625 cover. The front surface of the Inconel 625 cover was coated with a thin (~10-15 μm) layer of a high-temperature black coating made of Pyromark 2500 for laser photothermal absorption and thermal emission [26]. The back side of the Inconel 625 cover (in contact with the molten salt) was coated with a 200 nm thick Pt layer to avoid possible corrosion by the salts. Graphite gaskets, serving as mechanical seals, were inserted between the front and back plates.

## 2.2.Heat transfer model and data fitting

In the MPR measurement, the molten salt, Inconel 625 cover, and the Pyromark coating formed a three-layer structure (Figure 1b). The intensity-modulated continuous laser served as periodic heat flux and is applied on the top surface of the three-layer structure. Since the laser heating spot size (~15 mm) was much larger than the sample thickness (~3 mm) or the thermal penetration depth (< 2 mm) within the modulation frequency used in this study, we can assume the heat transfer is one-dimensional (1D). The 1D heat transfer model for a multi-layer sample with a heat flux imposed on the top surface is given by [26, 27],

$$\begin{bmatrix}\theta_b \\ q_b\end{bmatrix} = M_n M_{n-1} \cdots M_i \cdots M_1 \begin{bmatrix}\theta_s \\ q_s\end{bmatrix} = \begin{bmatrix}a(i\omega) & b(i\omega) \\ c(i\omega) & d(i\omega)\end{bmatrix}\begin{bmatrix}\theta_s \\ q_s\end{bmatrix} \quad (1)$$

where $\theta_s$ and $q_s$ are the surface temperature oscillation and the heat flux on the top surface, respectively. $\theta_b$ and $q_b$, on the other hand, are the temperature oscillation and the heat flux on the bottom layer, respectively. The transfer matrix $M_i$ is given by:

$$M_i = \begin{bmatrix}\cosh(D_i\sqrt{j\omega}) & -\frac{\sinh(D_i\sqrt{j\omega})}{e_i\sqrt{j\omega}} \\ -e_i\sqrt{j\omega}\sinh(D_i\sqrt{j\omega}) & \cosh(D_i\sqrt{j\omega})\end{bmatrix} \quad (2)$$

where $D_i = \frac{l_i}{\sqrt{\alpha_i}}$, $l_i$ and $\alpha_i$ are the thickness and the thermal diffusivity of the $i$ th layer respectively; $e_i$ is the thermal effusivity of the $i$ th layer, defined as $e_i = \sqrt{\rho_i c_{pi} k_i}$ , where $\rho_i$, $c_{pi}$and $k_i$ are the density, specific heat capacity and thermal conductivity of the $i$ th layer. In the current three-layer case, the 1st, 2 nd and, 3 rd layer are the Pyromark coating, Inconel 625 sheet, and the molten salt layer, respectively. $a(i\omega)$, $b(i\omega)$, $c(i\omega)$, and $d(i\omega)$ are the matrix elements of the lumped transfer matrix. With the adiabatic boundary conditions, equation (1) is simplified as:

$$q_b = c(i\omega)\theta_s + d(i\omega)q_s = 0 \tag{3}$$

At a low angular frequency, the thermal penetration depth ($\delta_p = \sqrt{\frac{2k}{\rho c_p \omega}}$) is long enough to probe into the molten salt layer. The thermal response from the sample is mainly determined by the thermophysical properties of the molten salt layer. When the angular frequency increases, the thermal penetration depth is shortened. It allows us to probe into the layers that are closer to the top surface, which corresponds to the Inconel 625 cover and the Pyromark coating. The thermal response from the sample is then determined by the thermophysical properties of the alloy and the coating.

The heat loss due to the radiative heat transfer from the laser window was considered in our model. The surface temperature $T_s$ at the laser window is the sum of a D.C. component and an A.C. component of the temperature, $T_s = T_{DC} + \theta_s$ , where $T_{DC}$ is the mean surface temperature. The radiation heat loss thus can be decomposed into the D.C. and A.C. components of radiation heat loss. The D.C. component is defined as

$$q_{loss,DC} = \varepsilon\sigma(T_{DC}^4 - T_0^4) \tag{4}$$

where $\varepsilon$ and $\sigma$ are the emissivity of the laser window and the Stefan–Boltzmann constant. $T_0$ is the ambient temperature (20 $^{o}$C). The A.C. component can be approximated as:

$$q_{loss,AC} \approx 4\varepsilon\sigma T_{DC}^3\theta_s \tag{5}$$

Considering the heat loss through thermal radiation, equation (3) is further modified by including the radiative heat flux:

$$c(i\omega)\theta_s + d(i\omega)(q_s - q_{loss,AC}) = 0 \tag{6}$$

An iterative process was introduced to extract the exact radiation heat loss, as shown in Figure 2a. Initially, temperature oscillation $\theta_{s,0}$ based on equation (3) with a known $q_s$ is applied to estimate radiation heat loss in equation (5). Considering the radiation heat loss at top surface, the net heat flux flowing into the multi-layer structure needs to be modified by subtracting radiation heat loss from $q_s$ at top surface, as instructed in equation (6). Based on equation (6), $\theta_{s,1}$ for the next round iteration can be estimated. Within each iteration step $i$, the difference in $\theta_{s,i}$ and $\theta_{s,i+1}$ between two consecutive steps is monitored. Once the difference is less than a tolerance, $e$, the radiation heat loss and temperature oscillation, $\theta_{s,end}$, in the last step will be considered as an accurate solution. The surface temperature oscillation amplitude with radiation heat loss can therefore be estimated by the iteration using equation (6).

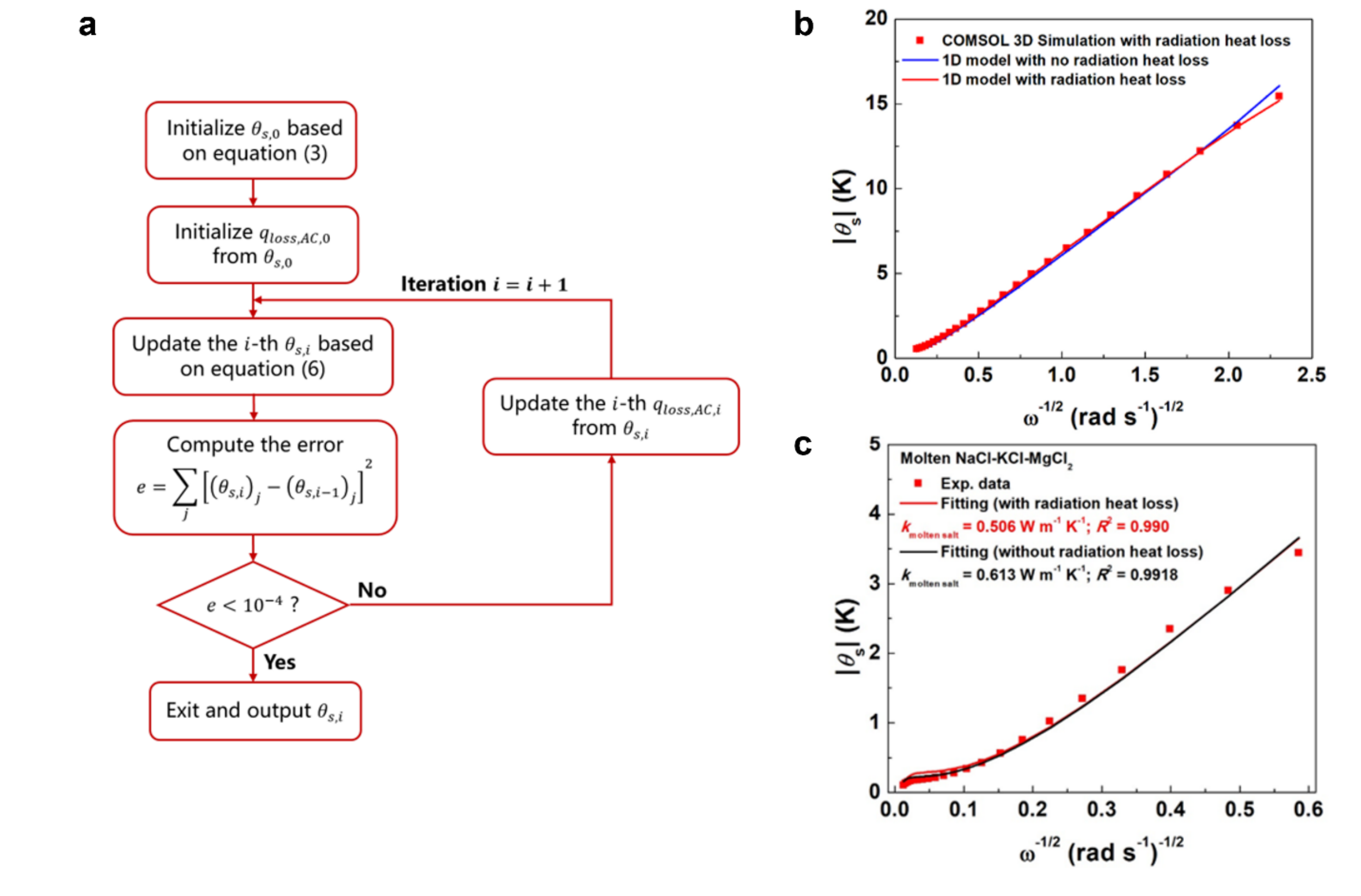

**Figure 2.** (**a**) Iterative process of radiation loss in 1D MPR analytical model. **(b)** COMSOL simulation of the MPR experimental data compared with the 1-D three-layer model fitting results with and without radiation heat loss, for a molten salt with its holder shown in Fig. 1b. **(c)** Fitting

to the MPR experimental data of $|\theta_s|$ *vs.* $\omega^{-\frac{1}{2}}$ of molten NaCl–KCl–$MgCl_2$ at 450 $^oC$. The model-fitted results with and without radiation heat loss are plotted alongside for comparison. If the radiation heat loss was not considered, the fitting result would considerably overestimate the molten salt thermal conductivity (0.613 vs. 0.506 W $m^{-1}$ $K^{-1}$).

Without considering the radiation heat loss, the fitting result using equation (3) would overestimate the best-fitted thermal conductivity of molten salts. To illustrate this, MPR experimental data results was simulated using a 3D finite element model FEM using COMSOL to compare the 1D model fitting results with and without radiation heat loss with a sample temperature of 450 $^oC$. Figure 2b shows that the 1D analytical model with the radiation heat loss have excellent agreement with the 3D FEM model, whereas the 1D analytical model without considering the radiation heat loss predicts a higher surface temperature rise at low frequency (to the right of the *x*-axis). This is because the radiation heat loss at this temperature is not negligible relative to the conduction heat transfer into the molten salt (a low thermal conductivity material) when the thermal penetration depth is long. The model fitting results of molten NaCl–KCl–$MgCl_2$ at 450 $^oC$ with and without considering the radiation heat loss were also compared (Figure 2c). The comparison shows that while both models can fit the data where, the model without considering the radiation heat loss overestimated the molten salt thermal conductivity by about 20 % (0.613 *vs.* 0.506 W $m^{-1}$ $K^{-1}$). Therefore, it can be concluded that radiation heat loss needs to be included in the model when measuring low thermal conductivity materials at high temperatures.

In this work, the MPR experimental data were fitted with the analytical model shown in Equation (6). Levenberg-Marquardt method was selected as the fitting method. It is a general approach for solving inverse heat transfer problems [30]. The thermal conductivity of molten salt ($k$) is the unknown to be estimated in this method and will be solved iteratively by an optimization algorithm. The thermophysical property of the 1st and the 2nd layers (i.e., Pyromark coating and Inconel 625), including $\rho$, $c_p$, and $k$ are the known input parameters. They can be found in refs [26, 31]. For the 3rd layer (i.e., paraffin wax, molten sulfur, and molten salts), $\rho$ and $c_p$ were the known input parameters from refs [32-35]. The thermal conductivity of molten salt $k$ was set as the fitting parameter. The aim of the method was to minimize the following least squares equation:

$$S = \sum_{i=1}^{N}(\theta_{s_{exp-i}} - \theta_{s_{sim-i}})^2 \tag{7}$$

where $\theta_{s_{exp-i}}$ is the measured surface temperature oscillation at a frequency point and $\theta_{s_{sim-i}}$ is the simulated surface temperature oscillation based on the known parameters and the initial guess value of the thermal conductivity. The iterative fitting algorithm is based on the following equation (8-9):

$$P^{n+1} = P^n + ((J^n)^{Tr}J^n + \mu^n\Omega^n)^{-1}(J^n)^{Tr}(\theta_{s_{exp}} - \theta_{s_{sim}}(P^n)) \tag{8}$$

$$\Omega^n = diag[(J^n)^{Tr}J^n] \tag{9}$$

where $P$ is the fitting parameters ($k$ of molten salt here), $J$ is the Jacobin Matrix of fitted parameters, $\mu$ is the damping factor, and $n$ indicates the iteration step. The iteration would stop and output the optimal value when the error $S$ is smaller than the specified error allowance, $1^{-4}$. More details about the iterative algorithm could be found elsewhere [36].

### 2.3.Sample preparation and loading

Two phase change materials (PCMs), namely, paraffin wax and sulfur, with established thermophysical properties were measured for calibration purpose. Paraffin wax (melting point of 53-58 °C, CAS no. 8002-74-2) was purchased from Sigma-Aldrich and sulfur (-325 mesh, purity ≥ 99.5 %, CAS no. 7704-34-9) from Spectrum. Three nitrate salts and one chloride salt were measured: $NaNO_3$, $KNO_3$, Solar salt ($NaNO_3$–$KNO_3$ mixture with wt.% 60–40), and NaCl–KCl–$MgCl_2$. $NaNO_3$ (purity ≥ 99 %, CAS reg. no. 7631-99-4) and $KNO_3$ (purity ≥ 99 %, CAS reg. no. 7757-79-1) were obtained from Sigma-Aldrich and used as the precursors of the solar salt. The solar salt was prepared by mixing the two pure nitrate salts based on the specified weight ratio. The salt mixture was melted and mixed repeatably in a clean glass crucible heated by a hot plate to ensure the uniformity of the composition. The entire procedure was performed inside a glovebox with < 0.5 ppm oxygen level. For the chloride salt, NaCl–KCl–$MgCl_2$ was prepared by NREL according to the published protocol (ref [33]). Table 1 summarizes the molten salts measured in this work.

**Table 1. Summary of molten salts measured in this work**

| Sample No. | Salt Composition | Reported melting point (°C) [32, 33] | Measured melting point (°C) |
|---|---|---|---|
| 1 | $NaNO_3$ (>99 %) | 308 | - |
| 2 | $KNO_3$ (>99 %) | 334 | - |
| 3 | $NaNO_3$–$KNO_3$ (wt.% 60–40) | 220 | 225 |
| 4 | NaCl–KCl–$MgCl_2$ (wt.% 15.11–38.91–45.98) | 401 | 401 |

Right before an MPR measurement, a solid salt was transferred from the glovebox and loaded to the MPR holder under ambident condition. The salt was then covered by the Pyromark-coated Inconel sheet, which was further anchored to the bottom plate with the top plate using screws and the graphite gaskets sealing. The holder was then connected with the vacuum pump and flowing Ar gas. The salt was purged with flowing Ar gas at a temperature about 50 $^{o}$C higher than its melting point for one hour to thoroughly remove the moisture that might have been introduced during the salt loading process. Afterward, the measurement started from slightly above the melting points with temperature increment of 25 or 50 $^{o}$C until the highest measurement temperature was reached. During the measurements, the salt-filled cavity was also continuously purged with the Ar gas. A gaseous pressure of 780 Torr, slightly above the atmospheric pressure, was maintained throughout the measurement.

The systematic errors could come from the uncertainty in the instruments (IR detector, lock-in amplifier, et), uncertainty of laser heat flux, the calibration curve converting the IR detection voltage to temperature. Specific to molten salt or other liquid measurements, there could be errors introduced by the natural convection effect. For the heat flux, we used a standard reference sample (borosilicate glass) to calibrate the heat flux, which is proportional to the thermal effusivity of the reference sample. The uncertainties due to the heat flux and the instruments were determined to be ~6 % in our earlier work [5, 6]. The natural convection effect could occur when the thermal penetration depth ($\delta_p$) is large enough, leading to a high Rayleigh number ($Ra$) and a Nusselt number ($Nu$) much higher than 1. Rayleigh number is defined as $Ra = \frac{\beta \Delta T g \delta_p^3}{\nu\alpha}$, where $\beta$ is the thermal expansion coefficient of the fluid, $\alpha$ is the thermal diffusivity of the fluid, $\nu$ is the kinematic viscosity of the fluid $g$ is the gravitational constant, $\Delta T$ is the characteristic temperature rise (AC amplitude) during the measurements. Here, we can modulate the frequency of the laser intensity to control the thermal penetration depth ($\delta_p = \sqrt{\frac{\alpha}{\pi f}}$) such that $Nu$ is close to one (i.e., $Nu \cong 1$) to have negligible natural convection effect. Table S1 in *Supplementary Information* (SI) shows the cutoff frequency for each fluid measured in this work.

Random error could be caused by optical misalignment for each measurement which could cause small deviation of the heat flux from the calibrated one as well as small deviation of the view factor from the sample surface to the IR detector. To minimize this error, we mounted the laser, sample holder, and IR detectors to fixed positions on an optical table. The random uncertainties were determined from the standard deviation of multiple measurements on the same sample, as shown in Table S2 in SI. The random error was around 7.40 %. Considering both the systematic and random uncertainties, the total uncertainty of the molten salt thermal conductivity measurement in our work was 9.40 %.

### 2.4.DSC measurement

While literature values of $c_p$ of molten salts were used to convert the thermal effusivity to thermal conductivity in the MPR measurements, the melting point and $c_p$ of the molten salt mixtures were still measured using a differential scanning calorimeter (DSC) (404 F3 Pegasus®, NETZSCH) to confirm the composition and characteristics of the mixtures. During the measurement, nitrogen gas ($N_2$) was used as the protective gas. $c_p$ of a sample was determined by using the ratio method [37]. A pure sapphire disc was used as the standard. It was noticed that the creeping effect of molten salts occurred when Pt crucibles were used in a DSC measurement [33, 38]. When graphite crucibles were used instead, there was no creeping effect observed. The heating rate of the DSC measurement was 20 K min$^{-1}$. After obtaining the DSC signals of baseline (no sample loading), standard (sapphire), and the actual molten sample samples, $c_p$ of the molten salt was calculated by using Proteus®, an analysis program provided by Netzsch. The average $c_\mathrm{p}$ values were obtained from three measurements, each of which was based on a new sample loading.

## 3. Results and Discussion

### 3.1.Calibration of MPR using standard molten materials

The reliability of the MPR measurement with the custom-made molten salt holder is verified by measuring known molten materials, including paraffin wax and sulfur. Figure 3a and Figure 3b show the typical raw data of $|\theta_s|$ *vs.* $\omega^{-\frac{1}{2}}$ of the MPR measurement of paraffin wax at 100 $^oC$ and sulfur at 250 $^oC$ respectively, with the data for sulfur at other temperatures shown in Supplementary Figure S2. The frequency range for the measurement is carefully selected to ensure the thermal penetration depth $\delta_p$ is long enough to probe into the molten samples while minimizing any possible effect from the natural convection in the molten material. The experimental data was fitted with the three-layer model with the radiative heat loss effect, i.e., equation (6), with the thermal conductivity values of the wax and sulfur shown next to their respective curves in Figure 3a and Figure 3b. For molten paraffin wax at 100 $^oC$, $k$ is determined to be 0.162 ± 0.012 W $m^{-1}$ $K^{-1}$,which is consistent with the average literature value of the same type of paraffin wax (0.160 ± 0.07 W $m^{-1}$ $K^{-1}$) [39, 40]. To demonstrate the effect of the natural convection on the measurements, paraffin wax was measured by scanning the sample within a low-frequency range where the natural convection effect is non-negligible. It found that at very-low-frequency (i.e., $f$ = 0.1 Hz), the natural convection in the molten paraffin wax layer is important, such that it can lead to a higher apparent thermal conductivity. The experimental results and the model fitting results for paraffin wax at very-low-frequency are included in Supplementary Figure S3.

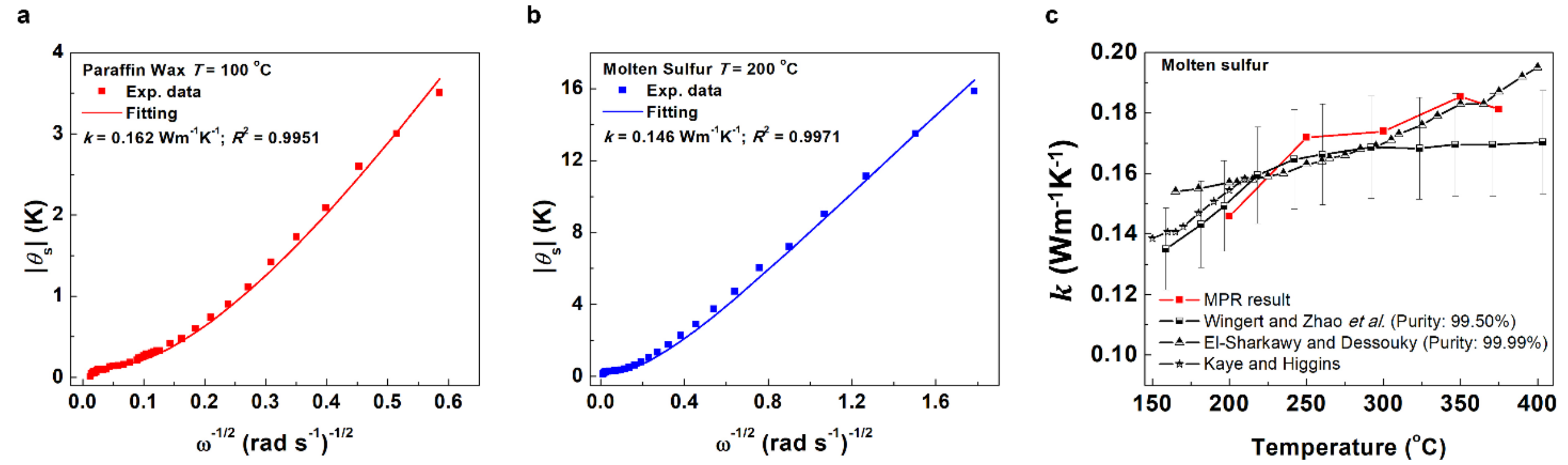

**Figure 3.** Plot of $|\theta_s|$ versus $\omega^{-\frac{1}{2}}$ plot of MPR measurements of **(a)** molten paraffin wax at 100 $^oC$, and **(b)** molten sulfur at 200 $^oC$ respectively. The fitted thermal conductivity are listed in the plots;

**(c)** thermal conductivity of molten sulfur as a function of temperature compared with the existing literature values [35, 39-42]: Wingert and Zhao *et al.* [43] (purity: 99.50 %), El-Sharkawy and Dessouky [35] (purity: 99.99 %), and Kaye and Higgins [42]

Figure 3c shows the measured $k$ of molten sulfur from 250 to 375 $^oC$. The average $k$ of molten sulfur increases from 0.146 to 0.181 W $m^{-1}$ $K^{-1}$ within the temperature range of 250 to 375 $^oC$. This result shows good agreement with the literature values of sulfur [35, 41, 42]. Our result is close to that reported by El-Sharkawy and Dessouky [35] in which the sulfur has a slightly higher purity (~99.9 %), and is only slightly higher than the results from the frequency-domain hot-wire method by Wingert and Zhao *et al.* [43], in which they used sulfur with a purity of 99.50 %. These calibration results demonstrate the reliability of the MPR technique for measuring molten materials.

### 3.2. Specific heat capacity measurements of molten salts

Figure 4 show the $c_p$ *vs.* $T$ curves for the solar salt mixture, $NaNO_3$–$KNO_3$ (wt.% 60–40), NaCl–KCl–$MgCl_2$ (wt.% 15.11–38.91–45.98. The same DSC measurements also yield the melting points of the salts, which are listed in Table 1 and are confirmed to be consistent with the literature values. The measured $c_p$ of $NaNO_3$–$KNO_3$ decreases from ~1.57 to 1.49 J $g^{-1}$ $K^{-1}$ when the temperature increases from 250 to 425 $^oC$. The results are consistent with the literature values [32] with a maximum difference of ~5 %. The measured $c_p$ of NaCl–KCl–$MgCl_2$ first decreases from ~1.12 to ~1.02 J $g^{-1}$ $K^{-1}$ from 450 to 500 $^oC$ and then keeps constant from 500 to 650 $^oC$. It is consistent with the values reported by NREL [33] as well as the theoretical prediction obtained using the software Factsage [33, 44]. In summary, our DSC measurements on the melting points and $c_p$ confirm that the composition and quality of our salt samples are similar to those used in the earlier studies on the same salts. The literature $c_p$ values of the selected molten salts are thus suitable to be used for the thermal conductivity calculation in the MPR measurements.

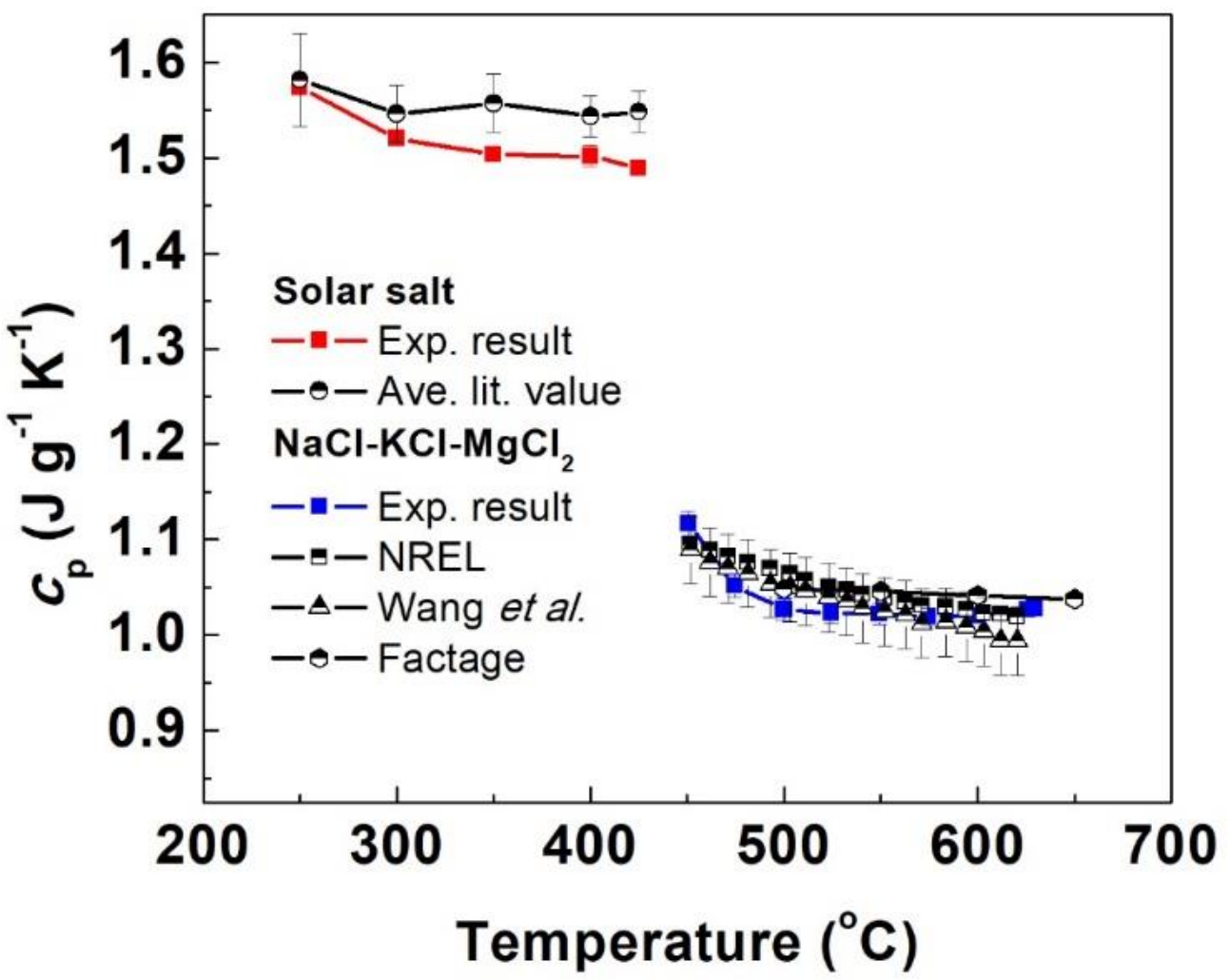

**Figure 4.** Measured specific heat capacity of **(a)** molten solar salt; and **(b)** molten NaCl–KCl–$MgCl_2$.The literature values are included in the plot for comparison. The reference data are from: D'Aguanno *et al.*, Wang *et al.* [33], NREL [33], and Factage [44]

### 3.3.Thermal conductivity of $NaNO_3$ and $KNO_3$

Figure 5 shows the measured temperature-dependent thermal conductivity of molten $NaNO_3$ and $KNO_3$. The low-frequency linear regions of the $|\theta_s|$ *vs.* $\omega^{-\frac{1}{2}}$ plots are shown in Supplementary Figure S4. Our MPR results show that the thermal conductivity of molten $NaNO_3$ slightly decreases from 0.562 to 0.540 W $m^{-1}$ $K^{-1}$ when the temperature increases from 350 to 450 $^oC$. For molten $KNO_3$, the measured thermal conductivity also shows a graduate decrease from 0.488 to 0.443 W $m^{-1}$ $K^{-1}$ from 350 to 450 $^oC$. Similar to the measurement of sulfur, our measured results of molten $NaNO_3$ and $KNO_3$ are close to that of Zhao *et al.* [12]. Both our MPR and Zhao *et al.*'s results are higher than most of the literature-reported values by 10-15 %, including those obtained from the LFA method. It is also noted that there is a large discrepancy among the literature-reported values. The data obtained by the steady state method might be problematic due to convection errors, the presence of radiation loss, and inaccurate heat flux calibration [11, 16, 45]; on the other hand, time-domain measurement data can be highly influenced by the time window selected for data processing. The time window has to be selected in order to eliminate errors from natural convection and the thermophysical properties of the sensor or confinement materials. Since the MPR and Zhao *et al.*'s methods are both based on frequency-domain techniques, which are expected to eliminate the convection effect and are less sensitive to the salt creeping effect, these results are likely better representatives of the true thermal conductivity of the two nitrate salts measured here. A detailed discussion on the differences between the MPR technique and the LFA method is included in *Section 3.6*.

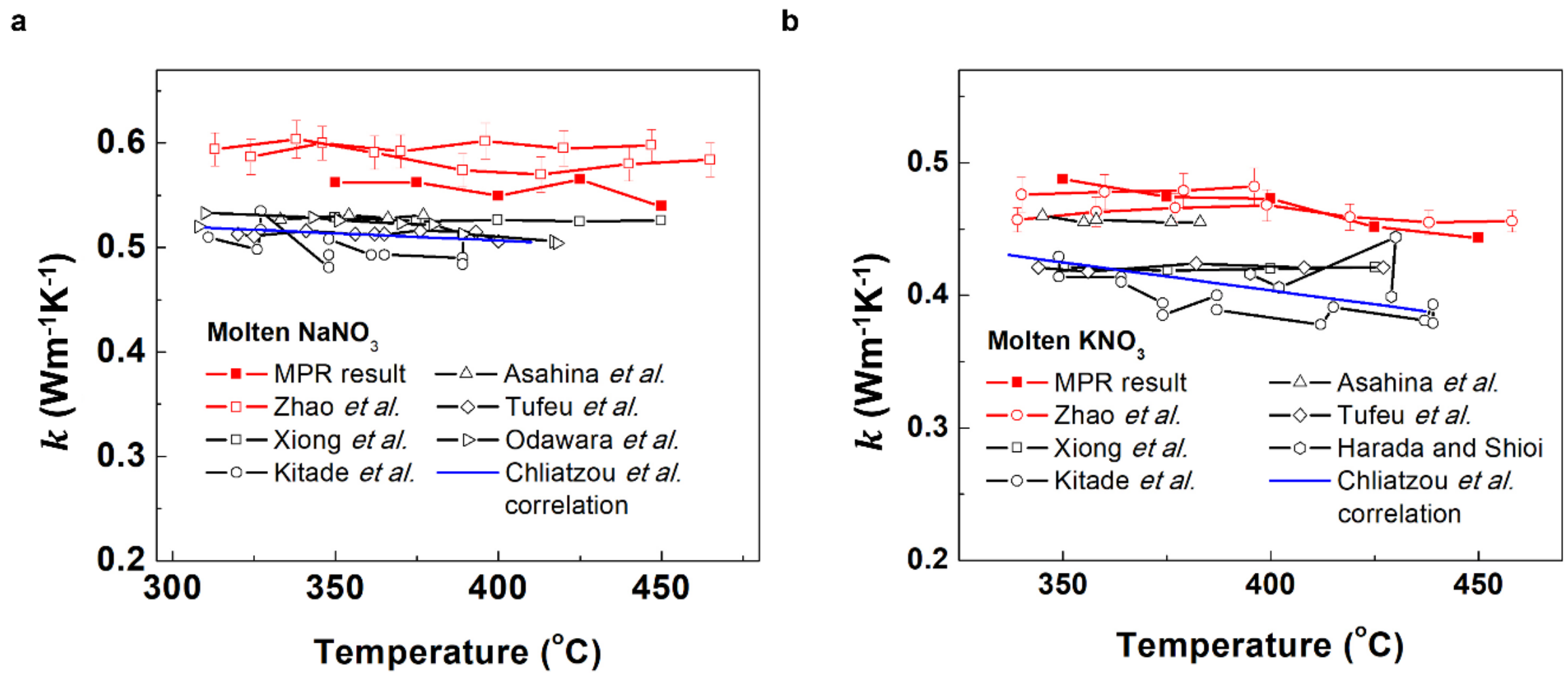

**Figure 5.** Measured temperature-dependent thermal conductivity of **(a)** molten $NaNO_3$ and **(b)** molten $KNO_3$, along with selected literature values. The reference data are from: Zhao *et al.* [12] (frequency-domain hot-wire method), Xiong *et al.* [46] (LFA), Kitade *et al.* [47] (THW), Asahina *et al.* [48] (pulse heated flat plate technique), Tufeu *et al.* [45] (steady state method), Odawara *et*

*al.* [49] (wave-front-shearing interferometry), Harada and Shioi [11] (LFA), and Chliatzou *et al.*'s correlation [11].

### 3.4. Thermal conductivity of Solar Salt ($NaNO_3$ -$KNO_3$ mixture)

Figure 6 shows the measured thermal conductivity of the molten solar salt. The experimental data of $|\theta_s|$ *vs.* $\omega^{-\frac{1}{2}}$ of this salt are shown in Supplementary Figure S5. The measured thermal conductivity of the molten solar salt slightly decreases from 0.510 to 0.477 W $m^{-1}$ $K^{-1}$ when the temperature increases from 300 to 425 $^{o}$C. Because thermal conductivity of the molten nitrate salt is mainly contributed by diffusons [25], or vibrational modes with approximately inter-molecular diffusion length, the thermal conductivity of the molten mixture is expected to lie in between those of its constituents, which is the case for our measured $k$ values.

Furthermore, the measured result is compared to the estimated one based on the rule of mixing, i.e.:

$$k_{\mathrm{NaNO3-\,KNO3}} = a k_{\mathrm{NaNO3}} + (1-a) k_{\mathrm{KNO3}} \tag{4}$$

where $k_{\mathrm{NaNO3}}$ and $k_{\mathrm{KNO3}}$ are the thermal conductivity of pure $NaNO_3$ and $KNO_3$ respectively from our MPR measurements (shown in Figure 5), and $a$ = 0.64 is the molar percentage of $NaNO_3$ salt. As shown in Figure 6, the measured $k$ value is about 10 % lower than that of the estimated one at 300 $^{o}$C but the two values are getting closer at higher temperatures. This result may be attributed to additional phonon scattering by different cations in the mixture, which suggests that the propagating phonons could still contribute to a small but non-negligible portion of thermal conductivity in the pure nitrate salts at around 300 $^{o}$C but this contribution is gradually diminished at higher temperature. Nevertheless, the mechanistic understanding of heat conduction in these salts would warrant further studies, perhaps using atomistic simulations [50-52].

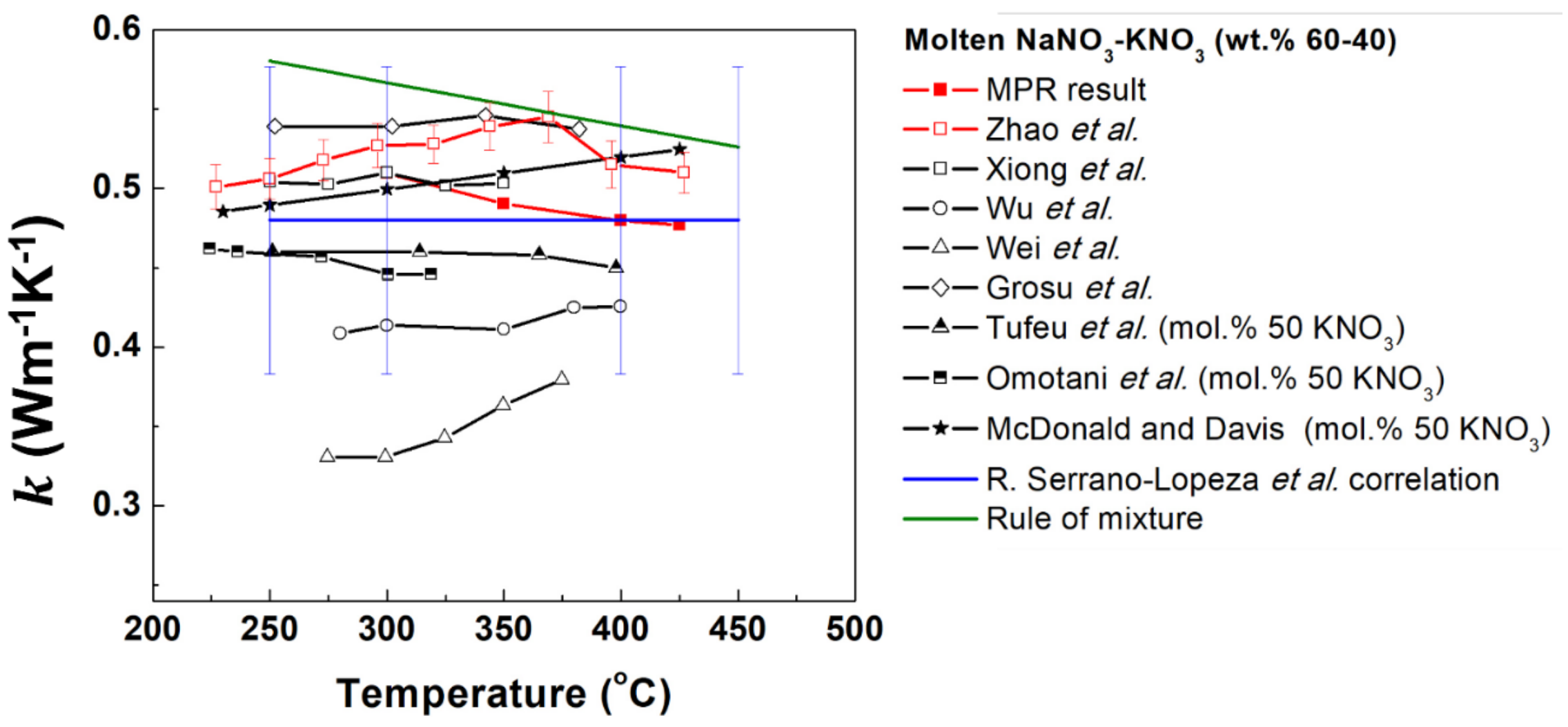

**Figure 6.** Thermal conductivity of molten solar salt along with selected literature results of $NaNO_3$–$KNO_3$ mixtures and the calculated thermal conductivity according to the rule of mixtures. The reference data are from: Zhao *et al.* [12] (frequency-domain hot-wire method), Xiong *et al.* [46] (LFA) , Wu *et al.* [53] (LFA), Wei *et al.* [54] (LFA), Grosu *et al.* [19] (LFA), Tufeu *et al.* [45] ($NaNO_3$–$KNO_3$ mol.% 50-50, steady state method), Omotani *et al.* [13] ($NaNO_3$–$KNO_3$ mol.% 50-50, THW), McDonald and Davis [16] ($NaNO_3$–$KNO_3$ mol.% 50-50, steady state method), and R. Serrano-López *et al.*’s correlation [55]

Figure 6 also compares the MPR results with the literature ones and the suggested correlations. Our measurement values are again in excellent agreement with those of Zhao *et al*. [12], as well as those by Gruso *et al.*[19] (using LFA with a Zn crucible) and Xiong *et al.* [46] (LFA). Our and Zhao *et al.*’s results are slightly higher than the correlation suggested by R. Serrano-López *et al.* (0.480 ± 0.098 W $m^{-1}$ $K^{-1}$) [55]. On the other hand, the LFA results from Wu *et al.* [53] and Wei *et al.* [54] show significantly lower values, even lower than that of pure $KNO_3$, which cannot be explained with the existing theories. On a similar nitrate salt mixture with 50 mol. % of $KNO_3$, the results from McDonald and Davis (steady state method) [16], Tufeu *et al.* (steady state method) [56], and Omotani *et al.* (THW) [13] also show discrepancies in both the absolute values and temperature trends. As mentioned in *Section 3.3*, the molten salt thermal conductivity obtained using the steady state method and time-domain measurement techniques can be susceptible to errors from convection, radiation loss, and biased time window for data processing. It reveals the fact that reliable molten salt thermal conductivity data are difficult to obtain using the steady state, THW, or LFA methods. A more detailed discussion of LFA measurements of molten salts is presented in *Section 3.6*.

### 3.5.Thermal conductivity of ternary chloride salts

Figure 7 shows the measured thermal conductivity of molten NaCl–KCl–$MgCl_2$ as a function of temperature. The available literature values are included for comparison. The $\rho$ and $c_p$ of each salt for the $k$ determination are extracted from ref [33] . The experimental data of $|\theta_s|$ versus $\omega^{-\frac{1}{2}}$ are included in Supplementary Figure S6. The measured thermal conductivity of molten NaCl–KCl–$MgCl_2$ slightly decreases from 0.506 to 0.400 W $m^{-1}$ $K^{-1}$ when the temperature increases from 450 to 650 $^oC$. The experimental result is generally consistent with the results reported by Wang *et al.* [33] using LFA on the same salt, although their measured thermal conductivity of this molten salt shows a stronger decreasing trend with temperature.

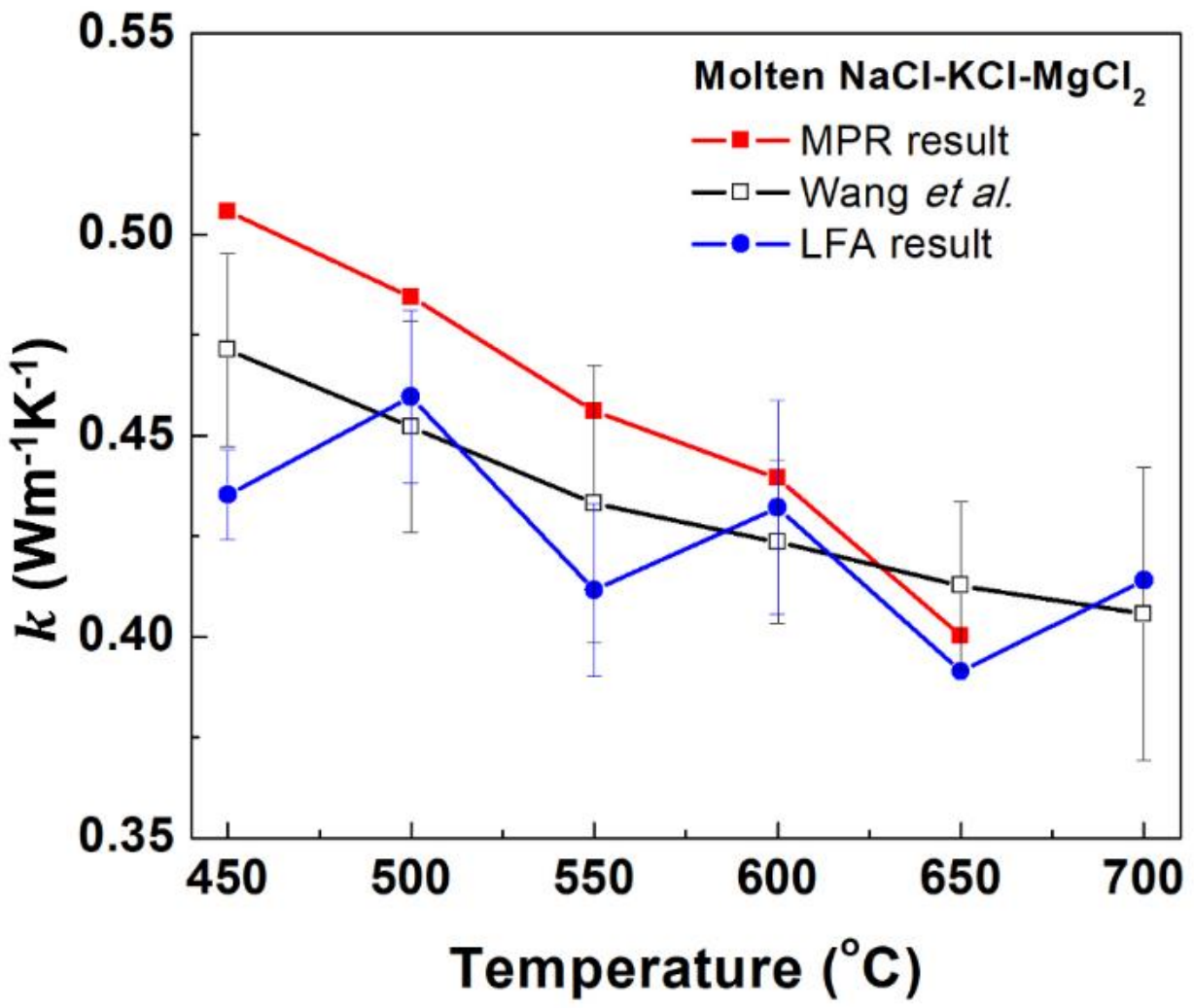

**Figure 7.** Measured temperature-dependent thermal conductivity of molten NaCl–KCl–$MgCl_2$. The reference data are from: Wang *et al.* [33] (LFA)

### 3.6. Comparison with other measurements

DiGuilio and Teja [57] have reviewed major thermal conductivity models for molten salts. They concluded that the thermal conductivity of molten salts should show a decreasing trend with increasing temperature. Recently, Zhao *et al.* [25] developed a phonon gas model for dense, strongly interacting liquids to predict the thermal conductivity of molten $NaNO_3$ and $KNO_3$. Their modeling results show that pure molten $NaNO_3$ and $KNO_3$ exhibit slightly decreasing trends in thermal conductivity as temperature increases. The modeling results match well with their previous reported experimental values using the frequency-domain hot-wire method [12]. According to our MPR experimental results, the measured thermal conductivity of molten salts in general shows a slightly decreasing or nearly constant temperature dependence. It is consistent with Zhao *et al.*'s model [25] and experimental data [12]. On the other hand, many of the thermal conductivity data of molten nitrate salts reported earlier show an increasing trend with temperature [58] (Figures 4

& 5). This could be attributed to the measurement errors originated from the stronger convection effect of molten salt and thermal radiation at higher temperature in conventional measurement techniques. Therefore, it is believed that the experimental results from the MPR technique and the frequency-domain hot-wire method are more reliable on molten nitrate salts, whereas other techniques such as LFA and the steady state method are more susceptible to larger errors as evidenced by the relatively large scattering of the data. Our MPR data are also in line with the recent LFA results on the two ternary chloride salts, suggesting that MPR is reliable on both nitrate and chloride salts while there could be special precautions needed for LFA on different types of molten salts, which warrants further investigation.

Here we specifically comment on several factors compounding the LFA because it has been a widely used technique in recent years. LFA is a time-domain method based on front side pulsed photothermal heating and back-side infrared thermometry (Supplementary Figure S7). According to discussion found in the literature and our own experience with the LFA, it was found that LFA measurement could be sensitive to various experimental conditions, such as the exact salt loading in the crucibles [19, 20, 59], type of crucibles used and their wettability with the molten salts (a non-wettable surface such as graphite could create a gap with the salt [19, 59, 60], whereas a wettable surface such as Pt could lead to significant salt creeping effect [19, 20]), the thickness of the salt layer in the crucibles (thicker layer could have significant natural convection effect [24, 59]), the overall measurement time (longer measurement duration could exacerbate the salt creeping effect [20]), exact temperature rise window used for data fitting [21-23], etc. The lack of a standard procedure of LFA for molten salts to control these factors could have contributed to the large scattering of the LFA data reported so far. Nevertheless, when properly done, the LFA method does yield reasonable results on the two chloride salts measured here. Consistent data on the NaCl–KCl–$MgCl_2$ salt was also obtained using our own LFA (Figure 6). Our experience as well as that of the authors of ref [33] shows that it is critical to minimize the salt creeping effect, e.g., by loading a suitable amount of salt and limiting the measurement time to about two temperature points for each salt loading.

The MPR method shows several advantages of addressing the issues in LFA (Figure 1). First, the MPR technique is a front side sensing method, which eliminates the influences of the convection effect or gaps between the salt and the backside of crucibles in LFA. Second, the MPR molten salt holder is vertically oriented whereas the LFA sample crucible is typically horizontally positioned. The vertical orientation of the holder avoids the issues due to the salt creeping effect because the salt will always be in contact with the front and back metallic sides of the holder due to their high wettability with the molten salts. The salt creeping in LFA could lead to non-uniform distribution of the horizontal molten salt layer inside the crucible. Lastly, MPR is a frequency-domain technique, so the thermal penetration depth can be controlled to be a few hundred micrometers within the molten salt volume, which minimizes the possible natural convection effect in the salt and heat losses from the crucible to the environment that could be present in a time-domain method. In LFA, typically only data within a certain time window are used for fitting to reduce the errors caused by the convection and heat loss effects [21-23]. However, there is no standard or quantitative guideline for the selection of the time window, which could appreciably impact the extracted thermal diffusivity.

## 4. Conclusion

The current paper presents the development of a non-contact, frequency-domain MPR technique as a convenient and reliable method of thermal conductivity measurement for molten salts. The MPR technique is calibrated using standard phase change materials, including paraffin wax and sulfur, demonstrating the suitability of the MPR technique in determining the thermal conductivity of molten materials. Nitrate and chloride salts for high-temperature CSP systems are measured using the MPR technique. Our results on $NaNO_3$, $KNO_3$, and solar salt show excellent agreement with those obtained by Zhao *et al*. [12] using another frequency-domain technique based on a hot-wire method. The MPR technique also yields reasonable results on chloride salts. The MPR technique avoids the complicated setup preparation procedures that are otherwise required in contact-based techniques. The frequency-domain feature of the MPR enables the precise control of the thermal penetration depth in the measurement, thus eliminating the possible convection heat transfer effect commonly encountered in the LFA and other conventional time-domain measurements. The use of the front side sensing and the vertical holder orientation also minimize the impact of the salt creeping effect, which is often an issue in LFA. The work demonstrated that the MPR technique developed here will be especially useful for measuring high temperature thermal conductivity of emerging molten salts and other challenging fluids (such as molten metals) for future thermal energy conversion and storage systems.

## Author statement

RC conceived the idea. KC and JZ designed the MPR setup. KC built the MPR setup. KC did the MPR measurements. TF did the modeling. TF and SA contributed to the LFA measurements. XZ contributed to the modeling. YZ, PL, YZ provided the salt samples. AZ and JEG provided the sulfur samples and contributed to the discussions. KC and RC wrote the manuscript, with input from all the coauthors. All the authors edited the manuscript.

## Supplementary Information

Photograph of the MPR experimental setup; Natural convection analysis for MPR measurement; Random error analysis; Experimental raw data of molten samples; Experimental details of the LFA method and the measurement results

## Acknowledgements

This material is based upon work supported by the U.S. Department of Energy's Office of Energy Efficiency and Renewable Energy (EERE) under Solar Energy Technologies Office (SETO) Agreement Number DE-EE0008379. The views expressed herein do not necessarily represent the views of the U.S. Department of Energy or the United States Government.

## Table List

### Table 1. Summary of molten salts measured in this work

| Sample No. | Salt Composition | Reported melting point (°C) [32, 33] | Measured melting point (°C) |
|---|---|---|---|
| 1 | $NaNO_3$ (>99%) | 308 | - |
| 2 | $KNO_3$ (>99%) | 334 | - |
| 3 | $NaNO_3$–$KNO_3$ (wt.% 60–40) | 220 | 225 |
| 4 | NaCl–KCl–$MgCl_2$ (wt.% 15.11–38.91–45.98) | 401 | 401 |

## Figures and Figure Captions

**Figure 1**

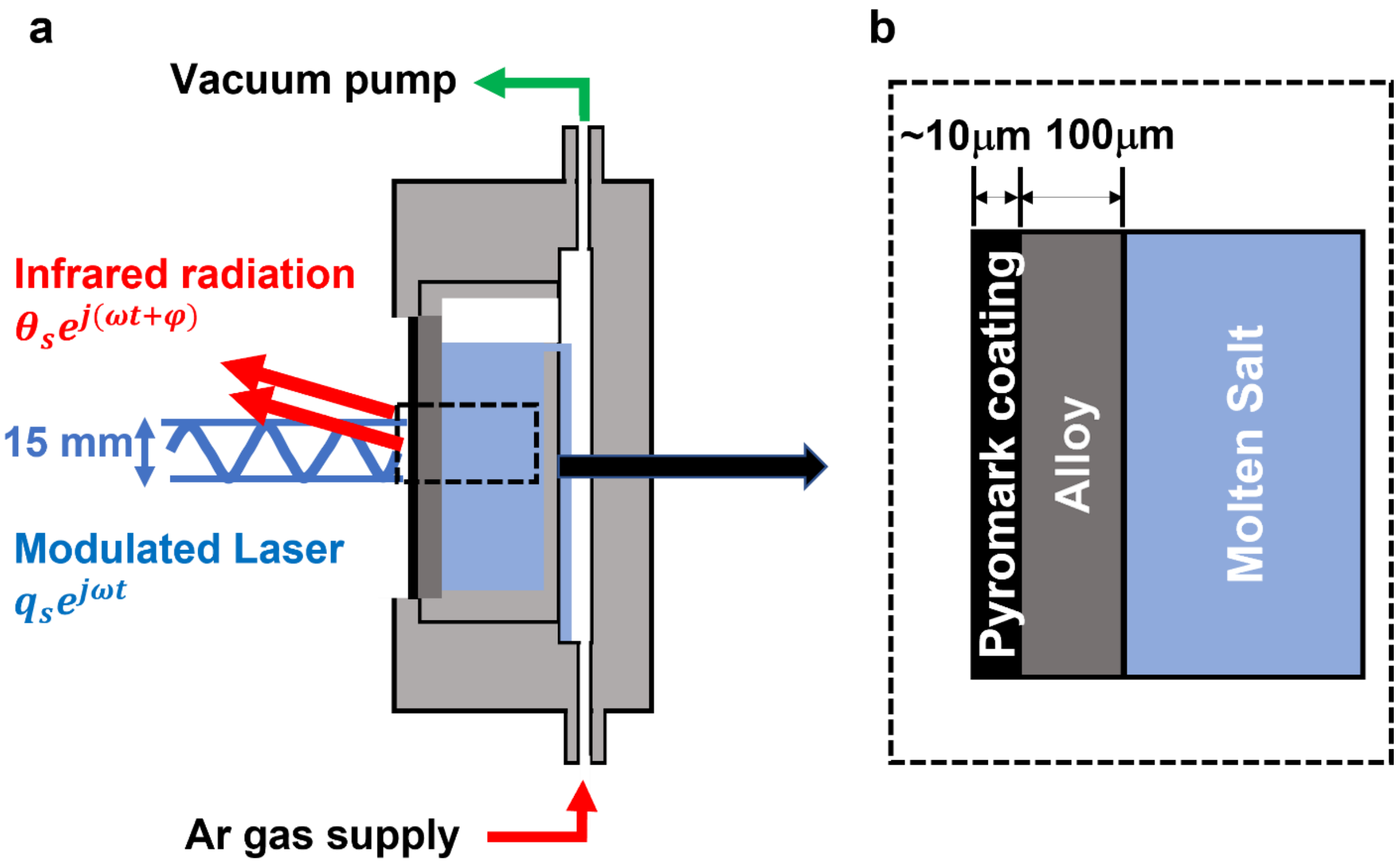

**Figure 1.** Schematics of the MPR molten salt measurement. **(a)** In the MPR measurement, the molten salt holder is vertically positioned. The thermal response from the front side of the molten salt holder is detected. The small cavity on the backside of the holder allows the overflow of the molten salt; **(b)** the three-layer structure formed by the molten salt, Inconel 625 cover and the Pyromark coating used for the heat transfer model in MPR.

## Figure 2

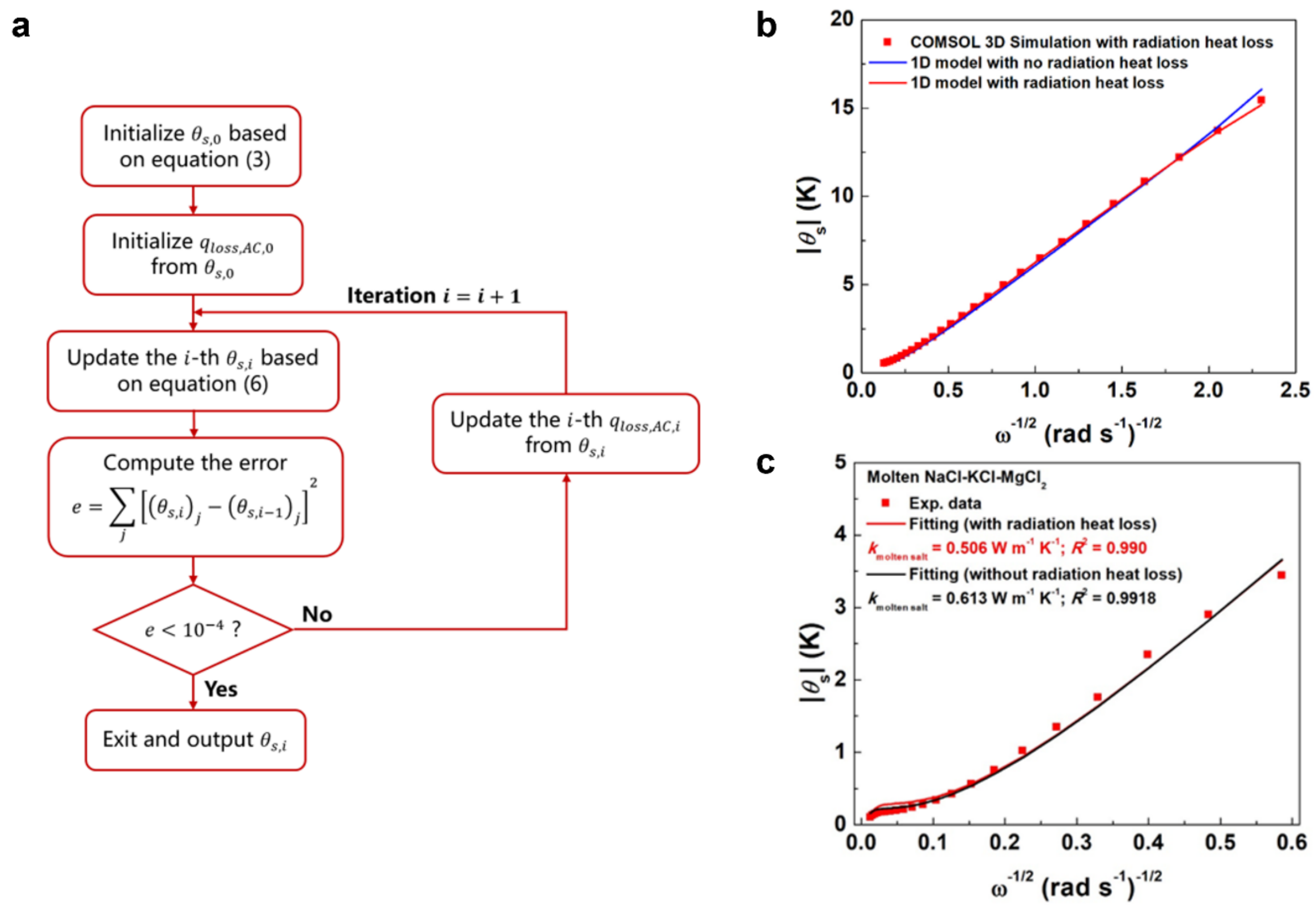

**Figure 2.** (**a**) Iterative process of radiation loss in 1D MPR analytical model. **(b)** COMSOL simulation of the MPR experimental data compared with the 1-D three-layer model fitting results with and without radiation heat loss, for a molten salt with its holder shown in Fig. 1b. **(c)** Fitting to the MPR experimental data of $|\theta_s|$ *vs.* $\omega^{-\frac{1}{2}}$ of molten NaCl–KCl–$MgCl_2$ at 450 $^{o}$C. The model-fitted results with and without radiation heat loss are plotted alongside for comparison. If the radiation heat loss was not considered, the fitting result would considerably overestimate the molten salt thermal conductivity (0.613 vs. 0.506 W $m^{-1}$ $K^{-1}$).

## Figure 3

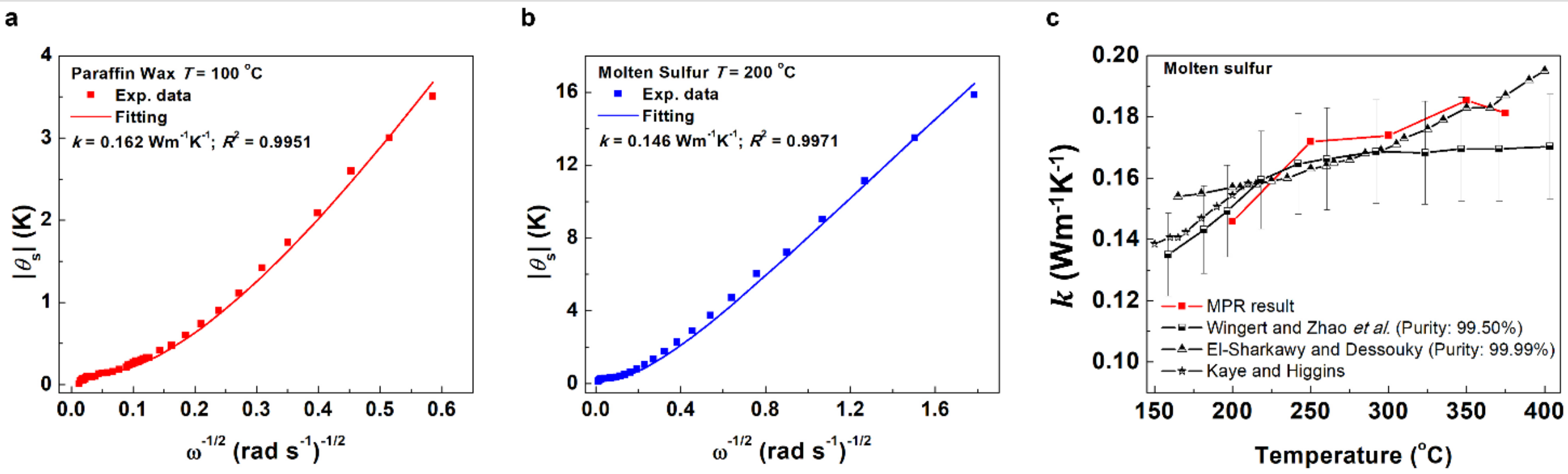

**Figure 3.** Plot of $|\theta_s|$ versus $\omega^{-\frac{1}{2}}$ plot of MPR measurements of **(a)** molten paraffin wax at 100 °C, and **(b)** molten sulfur at 200 °C respectively. The fitted thermal conductivity are listed in the plots; **(c)** thermal conductivity of molten sulfur as a function of temperature compared with the existing literature values [35, 39-42]: Wingert and Zhao *et al.* [43] (purity: 99.50%), El-Sharkawy and Dessouky [35] (purity: 99.99%), and Kaye and Higgins [42]

**Figure 4**

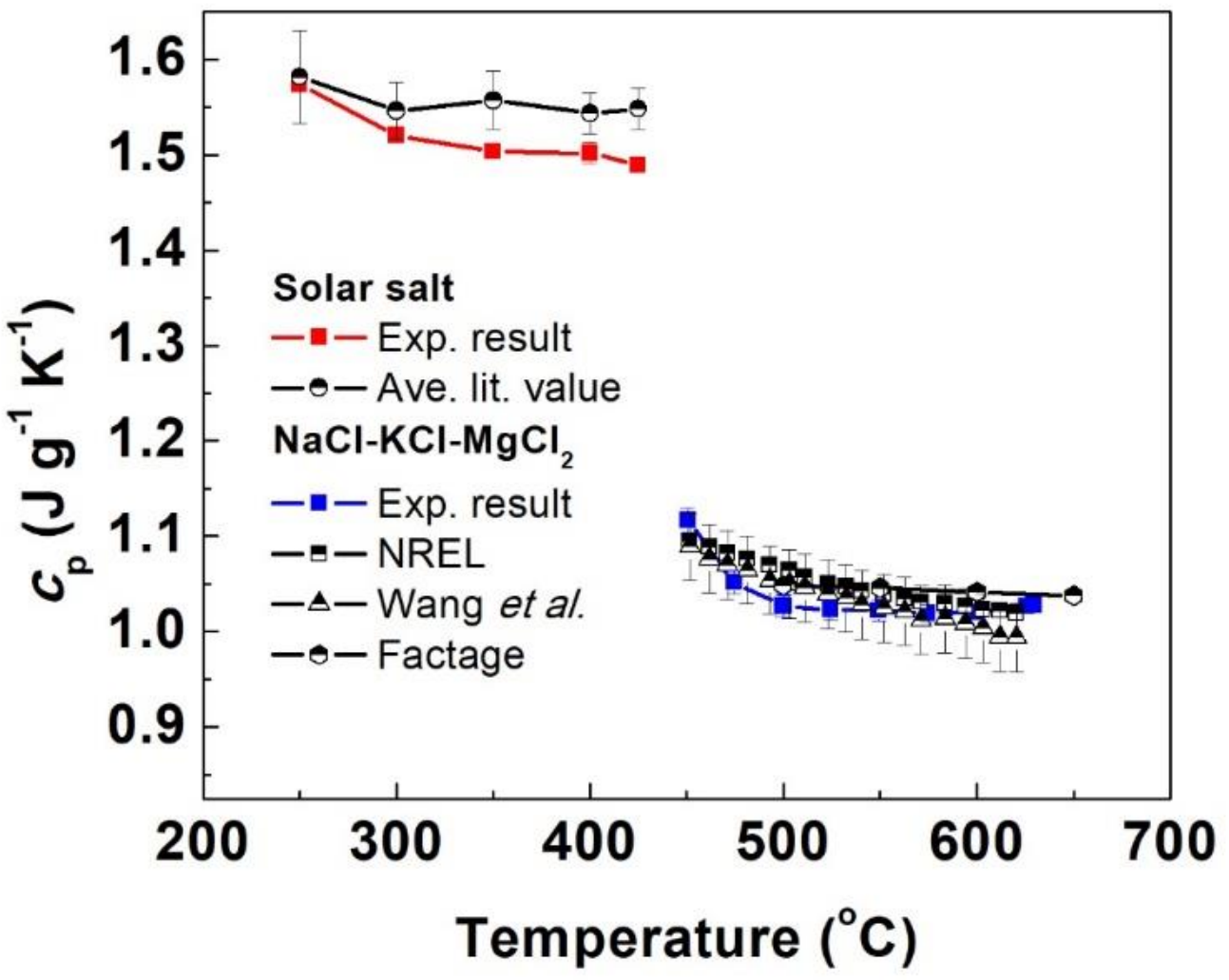

**Figure 4.** Measured specific heat capacity of **(a)** molten solar salt; and **(b)** molten NaCl–KCl–$MgCl_2$.The literature values are included in the plot for comparison. The reference data are from: D'Aguanno *et al.*, Wang *et al.* [33], NREL [33], and Factage [44]

**Figure 5**

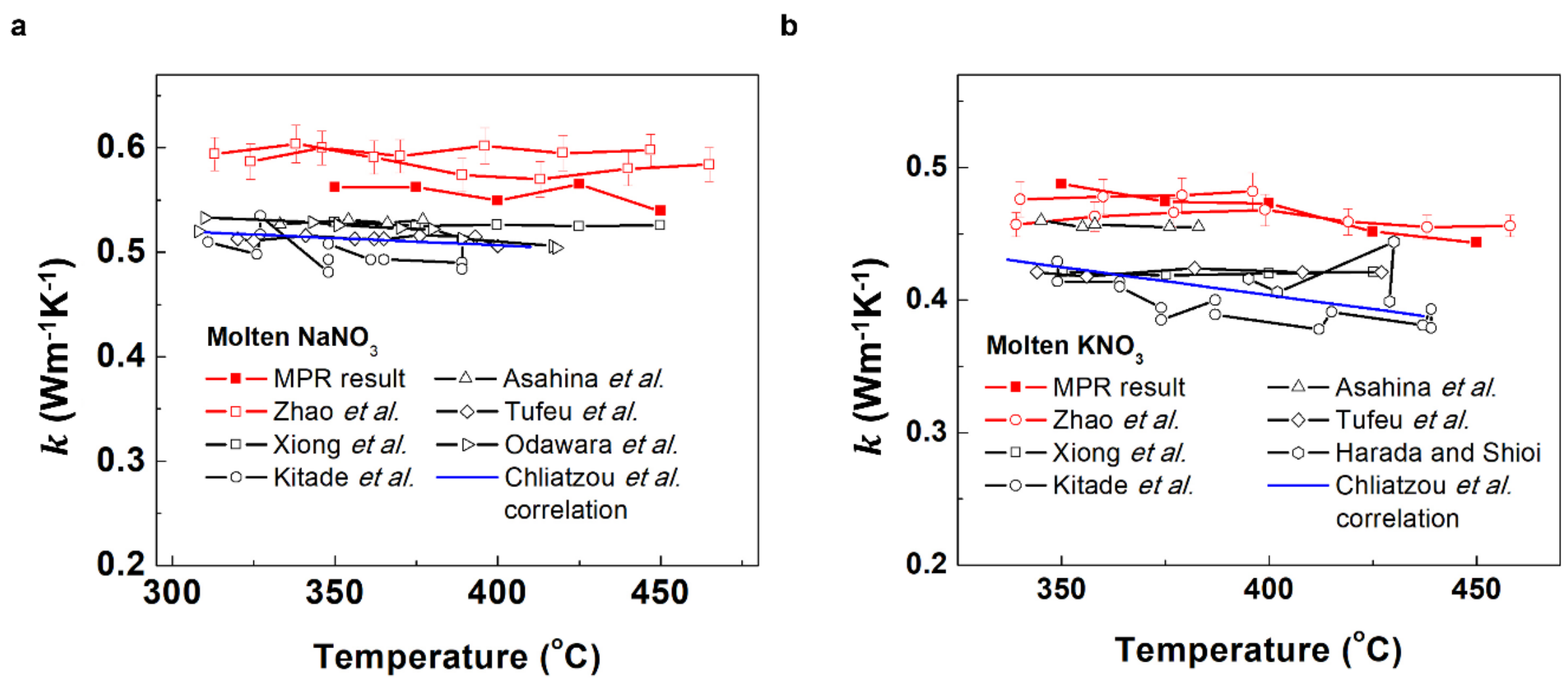

**Figure 5.** Measured temperature-dependent thermal conductivity of **(a)** molten $NaNO_3$ and **(b)** molten $KNO_3$, along with selected literature values. The reference data are from: Zhao *et al.* [12] (frequency-domain hot-wire method), Xiong *et al.* [46] (LFA), Kitade *et al.* [47] (THW), Asahina *et al.* [48] (pulse heated flat plate technique), Tufeu *et al.* [45] (steady state method), Odawara *et al.* [49] (wave-front-shearing interferometry), Harada and Shioi [11] (LFA), and Chliatzou *et al.*'s correlation [11].

**Figure 6**

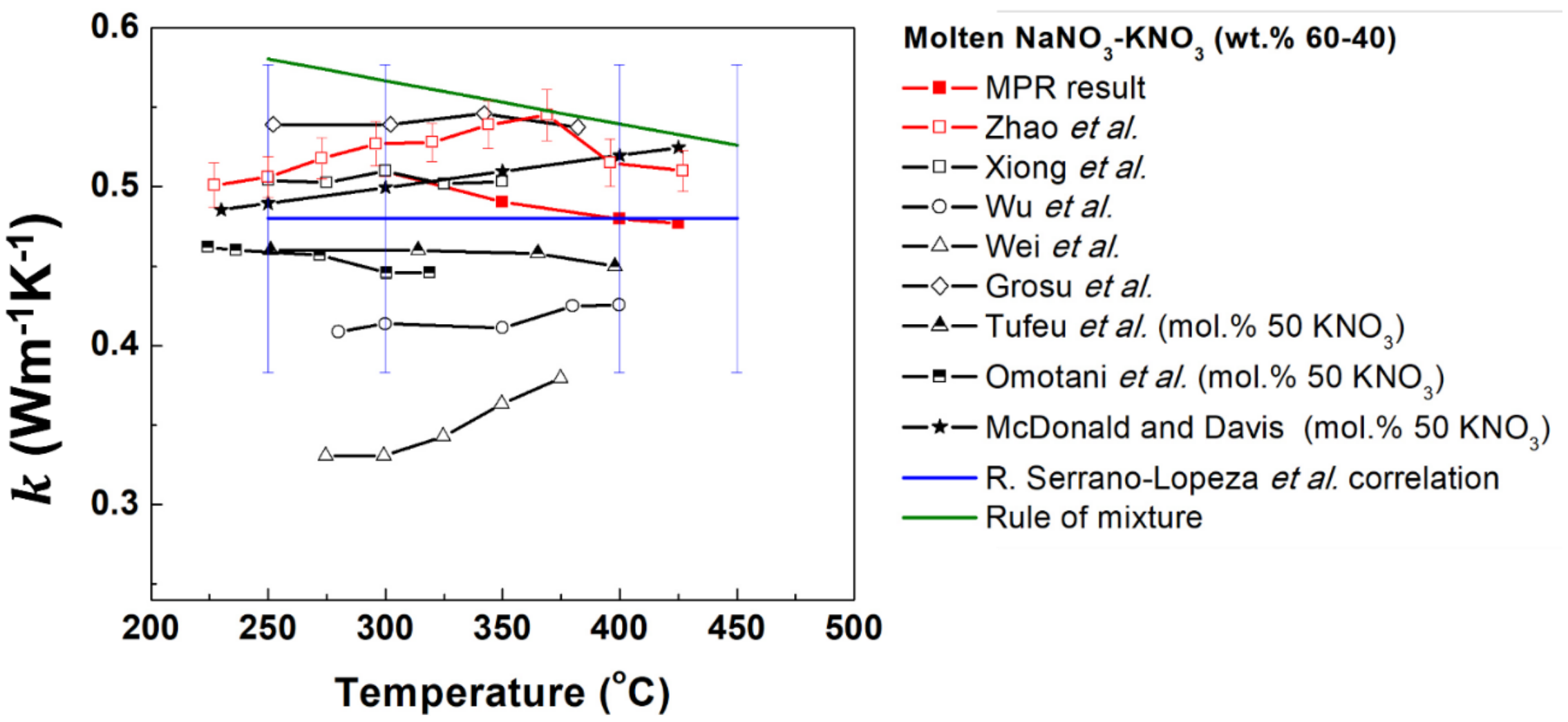

**Figure 6.** Thermal conductivity of molten solar salt along with selected literature results of $NaNO_3$–$KNO_3$ mixtures and the calculated thermal conductivity according to the rule of mixtures. The reference data are from: Zhao *et al.* [12] (frequency-domain hot-wire method), Xiong *et al.* [46] (LFA) , Wu *et al.* [53] (LFA), Wei *et al.* [54] (LFA), Grosu *et al.* [19] (LFA), Tufeu *et al.* [45] ($NaNO_3$–$KNO_3$ mol.%50-50, steady state method), Omotani *et al.* [13] ($NaNO_3$–$KNO_3$ mol.%50-50, THW), McDonald and Davis [16] ($NaNO_3$–$KNO_3$ mol.%50-50, steady state method), and R. Serrano-López *et al.*'s correlation [55]

**Figure 7**

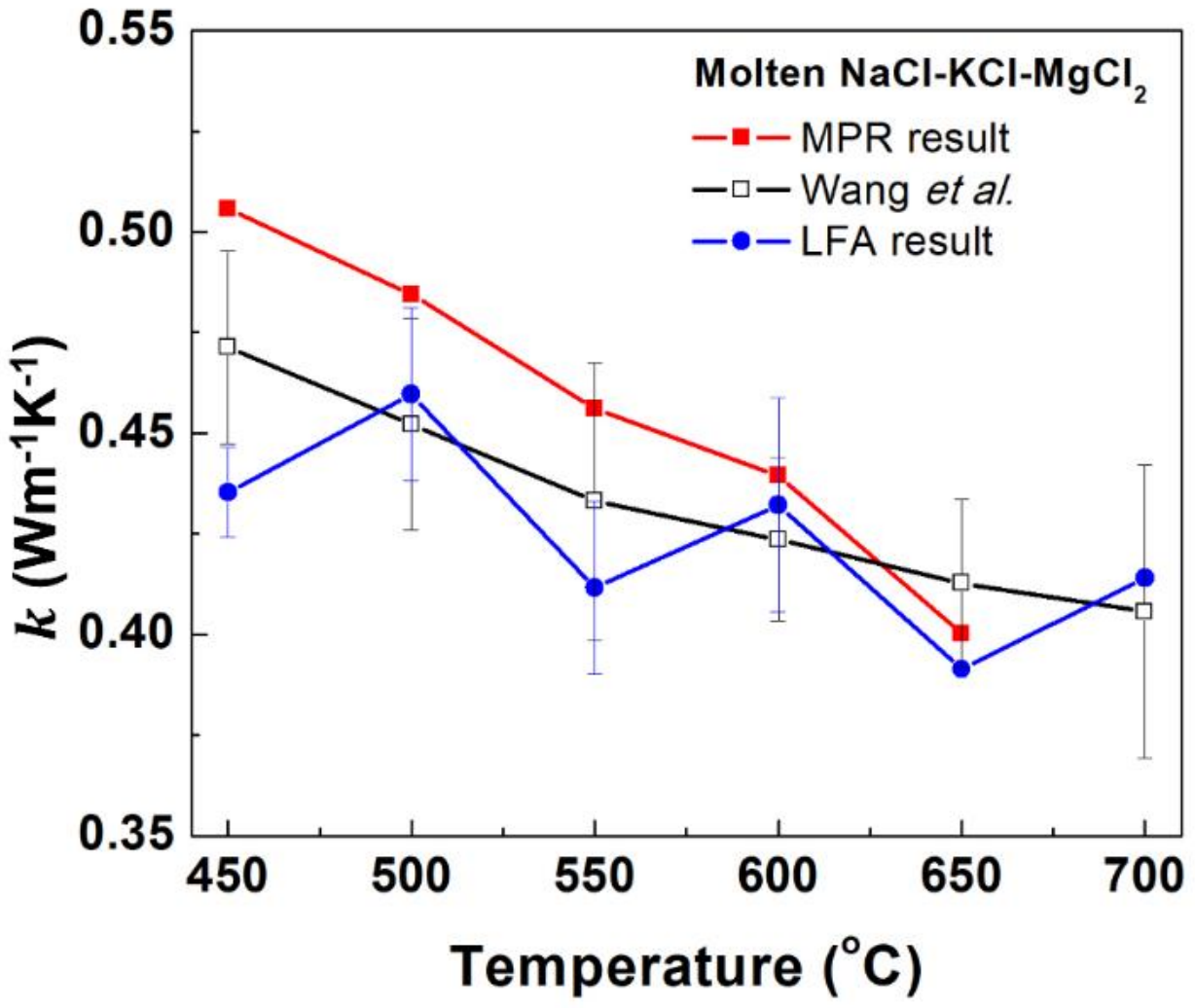

**Figure 7.** Measured temperature-dependent thermal conductivity of molten NaCl–KCl–$MgCl_2$.

The reference data are from: Wang *et al.* [33] (LFA)